\documentclass[aps, pra, noshowpacs,twocolumn,floatfix,superscriptaddress,amsmath]{revtex4-1}
\usepackage{graphicx}
\usepackage{graphics}
\usepackage{mathtools}
\usepackage{float}
\usepackage{amssymb}

\usepackage[dvipsnames]{xcolor}
\definecolor{darkblue}{rgb}{0.0, 0.0, 0.55}

\usepackage{color}
\usepackage[colorlinks=true,
            linkcolor=darkblue,
            urlcolor=darkblue,
            citecolor=darkblue]{hyperref}

\def \br{{\bf r}}

\def \mB{\mathrm{B}}

\def \mm{\mathrm{m}}
\def \mF{\mathrm{F}}

\def \mum{\mu \mathrm{m}}

\def \mmax{\mathrm{max}}

\def \tV{\tilde{V}_0}

\def \tVc{\tilde{V}_c}

\def \tVb{\tilde{V}}

\def \m2D{\mathrm{2D} }

\def \mms{\mathrm{ms}}

\def \mmax{\mathrm{max}}

\def \mb{\mathrm{b}}

\def \cD{{\cal{D}}}

\def \ka{\kappa}

\DeclareMathOperator\arctanh{arctanh}

\usepackage{xcolor}

\begin{document}
\title{Josephson junction dynamics in a two-dimensional ultracold Bose gas}
\author{Vijay Pal Singh}
\affiliation{Zentrum f\"ur Optische Quantentechnologien, Universit\"at Hamburg, 22761 Hamburg, Germany}
\affiliation{Institut f\"ur Laserphysik, Universit\"at Hamburg, 22761 Hamburg, Germany}
\affiliation{The Hamburg Centre for Ultrafast Imaging, Luruper Chaussee 149, Hamburg 22761, Germany}
\author{Niclas Luick}
\affiliation{Institut f\"ur Laserphysik, Universit\"at Hamburg, 22761 Hamburg, Germany}
\affiliation{The Hamburg Centre for Ultrafast Imaging, Luruper Chaussee 149, Hamburg 22761, Germany}
\author{Lennart Sobirey}
\affiliation{Institut f\"ur Laserphysik, Universit\"at Hamburg, 22761 Hamburg, Germany}
\affiliation{The Hamburg Centre for Ultrafast Imaging, Luruper Chaussee 149, Hamburg 22761, Germany}
\author{Ludwig Mathey}
\affiliation{Zentrum f\"ur Optische Quantentechnologien, Universit\"at Hamburg, 22761 Hamburg, Germany}
\affiliation{Institut f\"ur Laserphysik, Universit\"at Hamburg, 22761 Hamburg, Germany}
\affiliation{The Hamburg Centre for Ultrafast Imaging, Luruper Chaussee 149, Hamburg 22761, Germany}

\date{\today}
%
%
\begin{abstract}
We investigate the Berezinskii-Kosterlitz-Thouless (BKT) scaling of the critical current of Josephson junction dynamics across a barrier potential in a two-dimensional (2D) Bose gas, motivated by recent experiments by Luick \textit{et al.} arXiv:1908.09776. 
Using classical-field dynamics, we determine the dynamical regimes of this system, as a function of temperature and barrier height.  
As a central observable we determine the current-phase relation, as a defining property of these regimes. 
In addition to the ideal junction regime, we find a multimode regime, a second-harmonic regime, and an overdamped regime. 
For the ideal junction regime, we derive an analytical estimate for the critical current, which predicts the BKT scaling. We demonstrate this scaling behavior numerically for varying system sizes. 
The estimates of the critical current show excellent agreement with the numerical simulations and the experiments.
Furthermore, we show the damping of the supercurrent due to phonon excitations in the bulk, and the nucleation of vortex-antivortex pairs in the junction. 
\end{abstract}

\maketitle
%

\section{Introduction}

  For a Josephson junction created by a superconductor-insulator-superconductor interface, the Josephson relation $I_s = I_c \sin \phi$ relates the supercurrent $I_s$ to the phase difference $\phi$ of the order parameter  across the junction \cite{Josephson}.  $I_c$ is the critical current of the junction, which is the maximal supercurrent across the junction.
This connection is the defining functionality of quantum mechanical devices, such as superconducting quantum interference devices (SQUIDs).

 The on-going study of Josephson junctions (JJs) was broadened  in scope with the design of Josephson junctions in ultracold atom systems. 
This led to atomic JJs \cite{Inguscio, Oberthaler, Steinhauer, Thywissen, Betz, Inguscio2, Inguscio3, Schmiedmayer2018}, supercurrent dynamics in ring condensates \cite{Campbell2011, Campbell2014, Campbell2014N, Amy2014}, dc-SQUIDs  \cite{Ryu2013, CampbellSQ}, and quantum transport \cite{Ventra2015, Esslinger2017}.
Josephson tunneling between two condensates was studied in Ref.  \cite{Smerzi1997} and their theoretical investigation reported in Refs. \cite{Fantoni2000, Strinati2007, Dalfovo2007, Salasnich2009, Leggett1998, Ippei2005, Band2011, Leggett2001, Modugno2017, Zwerger1, Zaccanti, Marc}.   
Current phase relations of atomic JJs were measured in Refs. \cite{Niclas, Roati2019}. 
The decay of a supercurrent due to phase slip dynamics was discussed in Refs. \cite{Roati2018, RoatiAIP, Proukakis}. 
Temperature dependence of the phase coherence was measured in Ref. \cite{Oberthaler2006}.

Josephson junctions are utilized in phenomenological models of high-temperature superconductors, which describe these materials as stacks of two-dimensional (2D) systems coupled by JJs.  
Josephson junction arrays  also serve as a  model to describe transport phenomena in optically driven high-temperature superconductors \cite{Junichi}.
2D systems such as thin film superconductors \cite{ThinSC} or Josephson junction arrays \cite{Cuccoli}
undergo a Berezinskii-Kosterlitz-Thouless (BKT) transition within the XY universality class \cite{Berezinskii, Kosterlitz, Kosterlitz2}. 
The superfluid phase has quasi-long-range order and is characterized by a scale-invariant exponent $\tau$ of the single-particle correlation function $g_1$ that decays algebraically at large distances, $g_1(r) \sim |r|^{-\tau/4}$. 
At the transition the exponent assumes the critical value $\tau_c=1$ which is accompanied by a universal jump of the superfluid density.

   Recently, Ref. \cite{Niclas} reported on a study on Josephson junction dynamics of an ultracold 2D gas of $^{6}$Li atoms, which is realized by separating two uniform 2D clouds with a tunneling barrier. 
Using a strong barrier higher than the mean-field energy, the experiments measure the current-phase relation of an ideal junction, and the critical currents in the crossover from tightly bound molecules to weakly bound Cooper pairs.

  In this paper,  we establish a connection between the quasi-order scaling of 2D superfluids and the critical current of the Josephson junction.  
Specifically, we demonstrate the BKT scaling of the critical current of a Josephson junction coupling two 2D Bose gases using classical field simulations. 
The Josephson junction is created by a tunneling barrier between two 2D clouds of $^{6}\mathrm{Li}_2$ molecules, motivated by the experiments of Ref. \cite{Niclas}.
We find an interplay of the bulk and junction dynamics that is influenced by the barrier height and the temperature. 
Depending on these parameters, we find multimode (MM) and second-harmonic (SH) contributions to the current-phase relation.  
For large barrier heights we find that the junction dynamics displays ideal Josephson junction (IJJ) behavior, i.e. it  obeys the nonlinear current-phase relation $I(\phi)=I_c \sin(\phi)$.
We map out the MM, the SH, the IJJ, and an overdamped regime as a function of barrier height and temperature.
We determine the critical current numerically based on the current and phase dynamics at the barrier. 
The numerically obtained critical current, and an analytical estimate that we derive, show excellent agreement with the experimental values of Ref. \cite{Niclas}. 
We confirm the BKT scaling of the critical current by performing simulations for varying system sizes. 
The exponent of the critical current across the transition demonstrates agreement with the exponent of the corresponding equilibrium system, if the system is in the IJJ regime. 
Finally, we address the damping of the current and identify the damping mechanism which is due to phonon excitations in the bulk, and the nucleation of vortex-antivortex pairs in the junction.

This paper is organized as follows.
In Sec. \ref{sec:method} we describe our simulation method. 
In Sec. \ref{sec:bjj} we show the condensate dynamics and its dependence on the barrier height and the temperature. 
In Sec. \ref{sec:jc} we determine the critical current for an ideal junction and compare it to the simulation and the experiment. 
In Sec. \ref{sec:exp} we show the power-law scaling of the critical current  for varying system sizes. 
In Sec. \ref{sec:damping} we discuss the dissipation mechanism of the current, and in Sec. \ref{sec:con} we conclude.

\begin{figure*}
\includegraphics[width=1.00\linewidth]{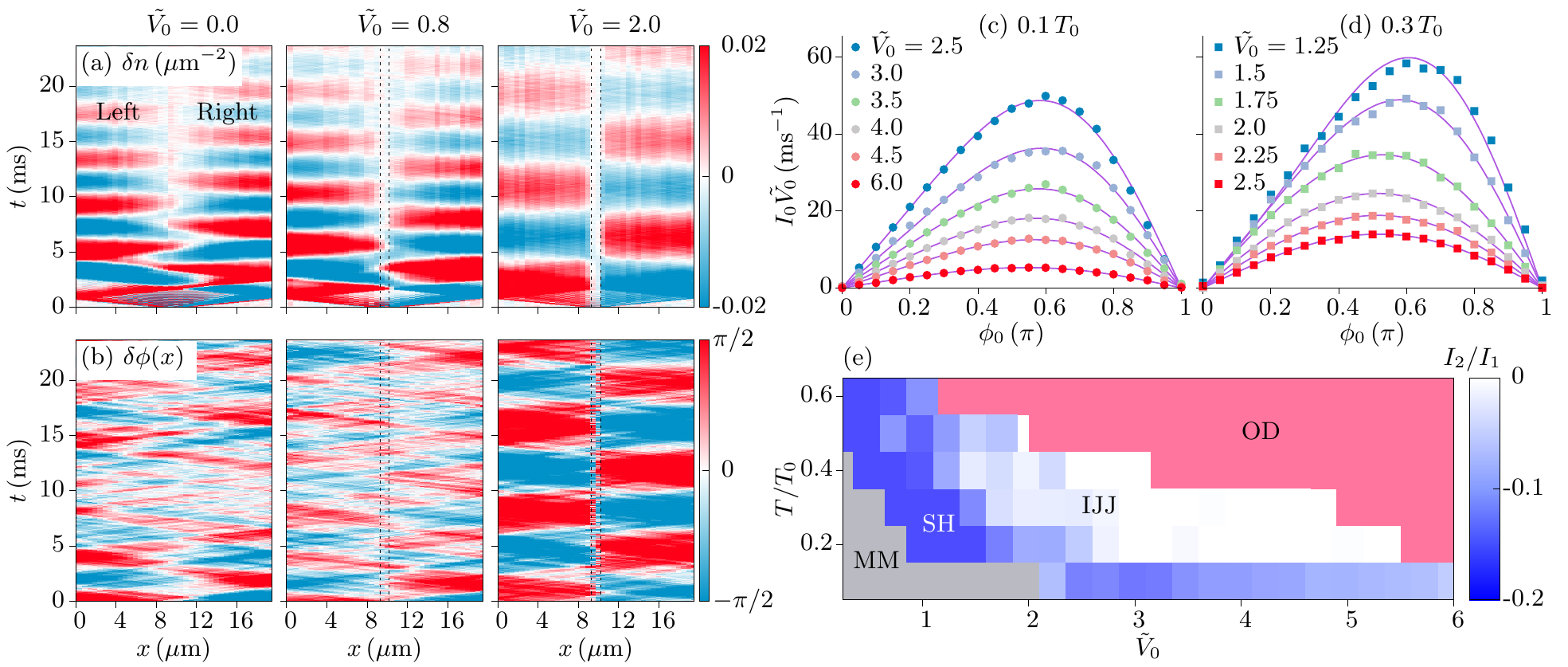}
\caption{\textbf{Dynamical regimes.} (a) Time evolution of the density $\delta n(x, t) = n(x,t) - n(x)$, which is averaged over the $y$ direction and the ensemble, for $\tilde{V}_0=0.0$, $0.8$, and $2.0$. $n(x)$ is the equilibrium density. 
Panel (b) shows the corresponding phase evolution $\delta \phi(x, t) = \phi(x,t) -  \phi_\mm (t)$ of a single trajectory of the ensemble.  
$ \phi_\mm$ is the mean phase.
We imprint a phase of $\phi_0=\pi/4$ on the left reservoir at $t=0$ for $n= 2.25\, \mum^{-2}$ and $T/T_0=0.3$. 
The barrier width $w$ is denoted by the two vertical dotted lines. 
Panels (c, d) show the current $I_0$ as a function of $\phi_0$ for various $\tilde{V}_0$ at $T/T_0=0.1$ and $0.3$, respectively.  
The continuous lines are fits with the fitting function $I(\phi_0) = I_1 \sin(\phi_0) + I_2 \sin(2\phi_0)$. 
(e) The dynamical regimes are multimode (MM), second-harmonic (SH), ideal Josephson junction (IJJ), and overdamped (OD), see text.   }
\label{Fig:dynamics}
\end{figure*}

\section{Simulation method}\label{sec:method}

 Motivated by the experiments \cite{Niclas} we study 2D clouds of  $^{6}\mathrm{Li}_2$ molecules confined in a box of dimensions $L_x \times L_y$. 
We simulate the dynamics using the c-field method of Ref. \cite{Singh2017}. 
The system is described by the Hamiltonian
\begin{equation} \label{eq_hamil}
\hat{H}_{0} = \int \mathrm{d}{\bf r} \Big[  \frac{\hbar^2}{2m} \nabla \hat{\psi}^\dagger({\bf r}) \cdot \nabla \hat{\psi}({\bf r})  + \frac{g}{2} \hat{\psi}^\dagger({\bf r})\hat{\psi}^\dagger({\bf r})\hat{\psi}({\bf r})\hat{\psi}({\bf r})\Big].
\end{equation}
$\hat{\psi}$ ($\hat{\psi}^\dagger$) is the bosonic annihilation (creation) operator. 
The interaction $g$ is given by $g=\tilde{g} \hbar^2/m$, where $\tilde{g}$ is the dimensionless interaction, and $m$ the molecular mass.
$\tilde{g}$ is determined by $\tilde{g}=\tilde{g}_0/\bigl(1- \frac{\tilde{g}_0}{2\pi} \ln(2.09 k_\mF \ell_z) \bigr)$, with  $\tilde{g}_0= \sqrt{8 \pi} a_s/\ell_z$ \cite{Turlapov2017}. $a_s$ is the molecular s-wave scattering length, $\ell_z= \sqrt{\hbar/(m \omega_z)}$ the harmonic oscillator length in the transverse direction, and $k_\mF$ the Fermi wavevector.
We discretize space on a lattice of size $N_x \times N_y$ and a discretization length $l=0.5\, \mu \mathrm{m}$. 
Within the c-field representation we replace the operators $\hat{\psi}$ in Eq. \ref{eq_hamil} and the equations of motion by complex numbers $\psi$. We sample the initial states in a grand canonical ensemble having chemical potential $\mu$ and temperature $T$ via a classical Metropolis algorithm. 
We choose the system parameters, such as the density $n$, $\tilde{g}$, and $T$ to be close to the experiments.  
In particular, we choose $n \approx 1.24\, \mu \mm^{-2}$, $T/T_0=0.3$, and $L_x \times L_y = 20 \times 40\, \mum^2$, which are the same as in Ref. \cite{Niclas}.  The critical temperature $T_0$ is estimated by $T_0 = 2\pi n \hbar^2/(m k_\mB \cD_c)$, where $\cD_c= \ln(380/\tilde{g})$ is the critical phase-space density \cite{Prokofev2001}. 
We vary $\tilde{g}$ in the range $1-4$ to cover the BEC regime of the experiments. 
For additional simulations, we  use $n \approx 2.25\, \mu \mm^{-2}$ and $\tilde{g}=1.8$, while we vary $T/T_0$ and the box size.

   To create the Josephson junction we add a barrier term $\mathcal{H}_{\mb}(t) = \int \mathrm{d}{\bf r}\, V({\bf r},t) n({\bf r})$, where $n({\bf r})$ is the density at the location  ${\bf r}=(x,y)$. The barrier potential $V({\bf r},t)$ is given by 
\begin{equation}\label{eq_pot} 
V({\bf r},t)  = V_0 (t) \exp \bigl(- 2 (x-x_0)^2/w^2 \bigr),
\end{equation}
where $V_0(t)$ is the time-dependent strength and $w$ the width. The potential is centered at $x_0= L_x/2$. 
We choose $w$ in the range $(0.85 - 2)\, \mum$ and $V_0$ in the range $ V_0/\mu \equiv \tilde{V}_0 = 0-6$, 
where $\mu = gn$ is the mean-field energy. 
We ramp up $V_0$ linearly over $150\, \mms$ and then wait for $50\, \mms$. 
This splits the system in $x$-direction into two uniform 2D clouds, which we refer to as the left and right reservoir. 
To create a phase difference across the junction, we imprint a phase  $\phi_0$ on the left reservoir, 
resulting in the phase difference $\phi_0= \phi_L - \phi_R$, where $\phi_L$ ($\phi_R$) is the mean phase of the left (right) reservoir.
The sudden imprint of the phase and the barrier lead to the dynamics displayed in Fig. \ref{Fig:dynamics}.   
We calculate the $x$ component of the  current density defined as: 
\begin{align}
j(x)= \frac{\hbar}{2imN_y} \sum_{y} ( \psi^\ast_{\br} \psi_{\br + l \hat{e}_x }  - \psi_{\br} \psi^\ast_{\br + l \hat{e}_x } ).
\end{align}
We calculate $j_1$ with $\br= (x_0  - l \hat{e}_x, y )$,  and $j_2$ with $\br= (x_0 ,  y )$.
$j= \langle j_1 + j_2 \rangle/2$ gives the averaged current density at the barrier center. 
This current fulfills the continuity equation of the density imbalance between the two reservoirs. 
The time evolution of the current is determined by $I(t) = j(t) L_y$, see Fig. \ref{Fig:damping}(a).  
We fit $I(t)$ with the function $f(t)=  I_0 e^{-\Gamma t} \sin(\omega t + \theta)$ to determine
the magnitude $I_0 \equiv |I_0| $, the damping rate $\Gamma$, the frequency $\omega$, and the phase shift $\theta$.

\section{Bulk versus junction dynamics}\label{sec:bjj}

    To characterize the junction dynamics we analyze the time evolution of the density and the phase of the system. 
As an illustration, we choose $n= 2.25\, \mum^{-2}$, $T/T_0=0.3$, and $\tilde{g}= 1.8$. 
We imprint a phase of $\pi/4$ on the left reservoir at $t=0$, for $\tV=0$, $0.8$, and $2$. 
We use $w=0.85\, \mum$, which in terms of the healing length $\xi$ is $w/\xi=2.4$, where $\xi =\hbar/\sqrt{2\mu m}=0.35\, \mum$.
In Fig. \ref{Fig:dynamics}(a) we show the time evolution of the density $\delta n (x, t)= n(x,t)- n(x)$, which is averaged over the y direction and the thermal ensemble. $n(x)$ is the averaged equilibrium density profile. 
For no barrier, $\tV=0$, the phase imprint creates two density pulses that are visible as density increase and decrease in the left and right reservoir, respectively. 
The density pulses propagate in opposite direction with the sound velocity and are reflected by the box edges.
For $\tV=0.8$, the barrier confines the density wave partially in each of the reservoirs.
For $\tV=2$, the density waves are well confined within the reservoirs as flow between the reservoirs is obstructed by the barrier. 
Instead, the density waves tunnel across the barrier, resulting in coherent Josephson oscillations between the reservoirs.

    In Fig. \ref{Fig:dynamics}(b) we show the time evolution of the phase $\delta \phi (x,t) = \phi(x,t) -  \phi_\mm (t)$ of a single trajectory, for the same $\tV$ as in Fig. \ref{Fig:dynamics}(a). 
$ \phi_\mm$ is the mean global phase.   
The sudden imprint of phase adds the mean phase difference $\phi_0 = \phi_L - \phi_R $ between the left and right reservoir at $t=0$. 
During the time evolution a phase gradient develops within the reservoirs, which results in $\phi_{L/R}$ being different from the phases close to the junction. 
For $\tV=0$, the phase evolves linearly with the distance. 
As the barrier height is increased, the phase gradient within the reservoirs decreases.
For $\tV=2$, the phase gradient within the reservoir almost vanishes, resulting in 
$\phi_{L/R}$ being the same as the phases in direct vicinity of the junction.   
This corresponds to IJJ dynamics.

  We now examine the current-phase relation (CPR) of the junction. 
We determine the current $I_0$ by fitting the time evolution of the current $I(t)$ to a damped sinusoidal function as described in Sec. \ref{sec:method}.
To obtain the CPR we calculate $I_0$ as a function of $\phi_0$. 
In Fig. \ref{Fig:dynamics}(c) we show  $I_0(\phi_0)$ for various values of $\tV$ at  $T/T_0 =0.1$,  and in Fig. \ref{Fig:dynamics}(d)  at $T/T_0=0.3$. 
We analyze these CPR curves by fitting them with a multi-harmonic fitting function  $I(\phi_0) = \sum_{n=1}^{n_\mmax} I_n \sin(n \phi_0) $, where we choose $n_\mmax = 5$.
We find that the result of these fits can be grouped into three regimes, as a function of the barrier height $\tV$ and the temperature. 
If the coefficient $I_1$ is dominant, the CPR reduces to the form of an IJJ, $I(\phi_0) =  I_1 \sin( \phi_0)  $.
If both $I_1$ and $I_2$ are non-negligible, we refer to the regime as second harmonic (SH) regime. 
In Fig. \ref{Fig:dynamics}(e) we indicate the crossover from IJJ to SH by depicting the ratio $I_2/I_1$. 
For lower temperatures and barriers, higher harmonical contributions become important. 
We indicate the multimode (MM) regime if $\sum_{n>2} (I_n/I_1)^2  > 0.02$, see also Appendix \ref{ap:sec:cpr}.
We note that CPR deviations were pointed out  for $\tV \lesssim 1$ by Refs. \cite{Strinati2007, Piazza2010, Watanabe2009} .
Furthermore, we indicate the overdamped (OD) regime based on the analysis shown in Sec.  \ref{sec:damping}.

  The results depicted in  Fig. \ref{Fig:dynamics}(e) demonstrate that the IJJ regime is strongly sensitive to the temperature and the barrier height. 
 This derives from the properties of the dynamical evolution shown in Figs.  \ref{Fig:dynamics}(a) and (b).
 The initial phase imprint creates phonon pulses in the two reservoirs. 
 For low temperatures and small barrier heights these pulses are weakly damped, which leads to the multimode regime. 
 For increasing temperature and barrier height, fewer and fewer of the phonon modes of the system contribute. 
 The increasing barrier height leads to a long tunneling time, which exceeds the damping time of more and more phonon modes, until the phase dynamics reduces to the dynamics of two global phases for each reservoir, as visible in Figs. \ref{Fig:dynamics}(a) and (b). 
 However, if the barrier is increased further, eventually the dynamics become overdamped. Here, the two reservoirs dephase on the timescale of the tunneling rate.   
We note that  the temperature dependence of the current-phase relation was measured in Nb/InAs/Nb junctions by Ref. \cite{Ebel2002} and in a weak link of $^4$He superfluids by Ref. \cite{Packard2006}.

\section{Critical current}\label{sec:jc}

\begin{figure}
\includegraphics[width=1.00\linewidth]{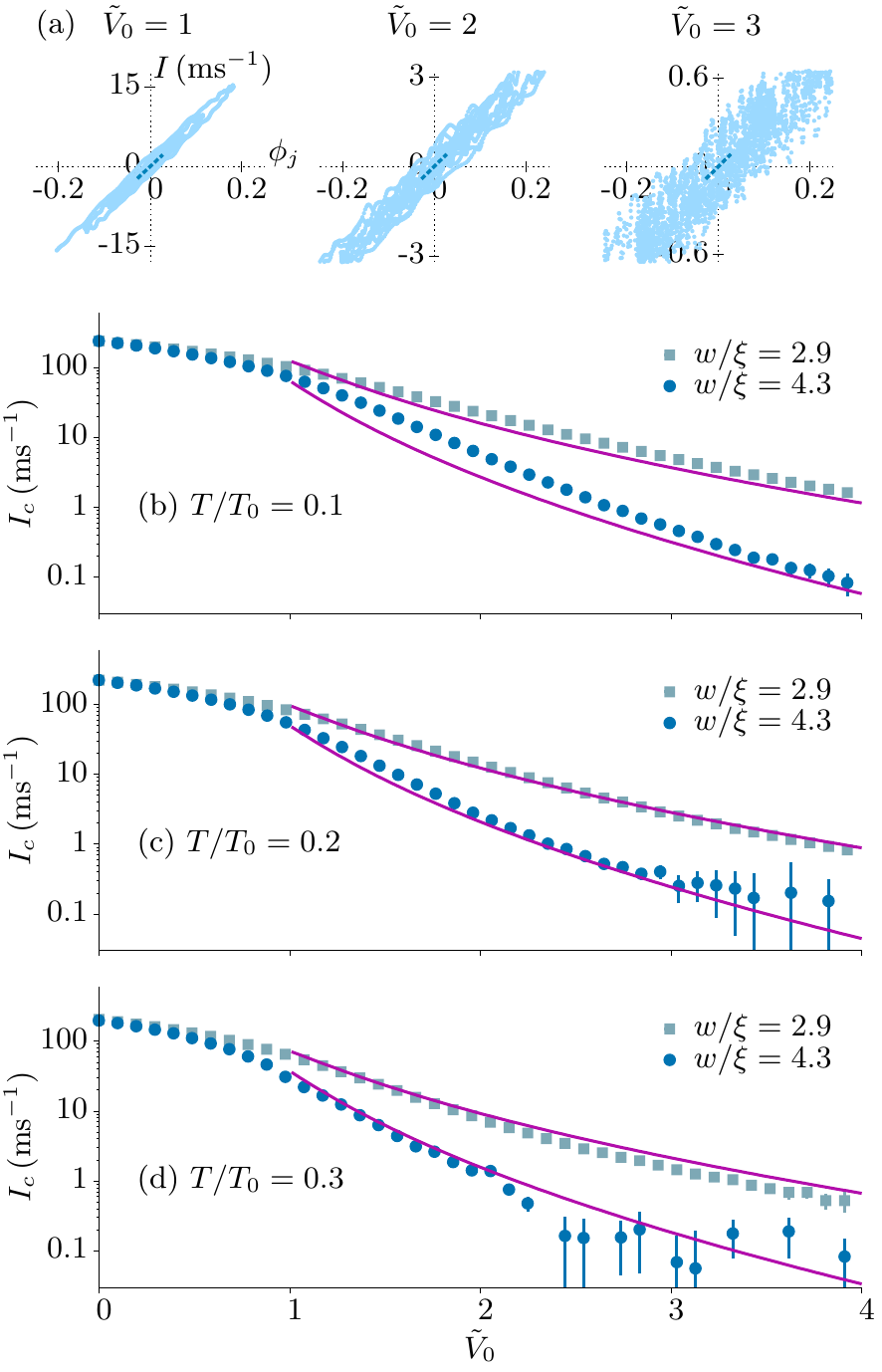}
\caption{\textbf{Critical current.} (a) Time evolution of the junction current $I(t)$ versus phase $\phi_j(t)$ for $\tV=1$, $2$, and $3$, depicted in the $I - \phi_j$ plane. 
The critical current $I_c$ is determined by the linear slope (dashed lines) of $I/\sin \phi_j$. 
Panels (b-d) show $I_c (\tV)$ for $T/T_0= 0.1$, $0.2$, and $0.3$, respectively, for $w/\xi=2.9$ (squares) and $4.3$ (circles).
The continuous lines are the estimates of Eq. \ref{eq:jc}.  
}
\label{Fig:Ic}
\end{figure}

  To determine the Josephson critical current, we calculate the junction phase  $\phi_j  \equiv  \langle  \phi_j  \rangle = 
 \langle \phi_{2, \mm} - \phi_{1, \mm}  \rangle$, where  $ \phi_{1/2, \mm} = \phi_{x_0 \mp 2 l \hat{e}_x } $ are the mean phases calculated by taking an average of the fields over the $y$ direction. 
We use the same density $n$ as above, and $T/T_0=0.2$. 
We calculate the time evolution of the current $I(t)$ and the phase $\phi_j(t)$.
In Fig. \ref{Fig:Ic}(a) we show the time evolution of $I(\phi_j)$ for $\tV=1$, $2$, and $3$.
In this small $\phi_j$ regime a linear behavior is observed. 
The width of the distribution increases with increasing $\tV$ due to increased phase fluctuations across the barrier.
We determine the critical current $I_c$ by the slope $I/\sin \phi_j$.  
Within the IJJ regime, this coincides with $I_1$.
$I_c$ decreases with increasing $\tV$, with $I_c = 83.8$, $15$, and $2.9 \, \mms^{-1}$ for $\tV=1$, $2$, and $3$, respectively.       
In Figs. \ref{Fig:Ic}(b)-(d) we show $I_c(\tV)$ for $T/T_0 = 0.1$, $0.2$, and $0.3$, and the barrier widths $w/\xi=2.9$ and $4.3$. 
As expected, $I_c$ is smaller for wider barriers. 
$I_c$ also decreases with the temperature as we show below.

   We derive an analytical estimate of the critical current by solving the mean-field equation and by considering single-particle tunneling across a rectangular barrier of height $V > \mu$ and width $d$, see Appendix \ref{ap:sec:Ic}. 
We obtain the current density 
\begin{align}\label{eq:jc:phij}
j = j_c \sin \phi_j,
\end{align}
with 
\begin{align}\label{eq:jc}
j_c = 2 n_0 \frac{\mu}{V + \sqrt{V^2 - \mu^2/2}}  \frac{ \hbar \ka }{m} \exp(-\ka d)
\end{align}
and
\begin{align}\label{eq:ka}
\ka^2 = \frac{m}{\hbar^2} \biggl( 3V - 2 \mu - \sqrt{V^2 - \mu^2/2} \biggr). 
\end{align}
$n_0$ is the condensate density. $\kappa$ is the damping parameter of the exponential wavefunction inside the barrier, 
which is determined variationally, and includes the mean-field repulsion under the barrier. 
We interpret this result for $j_c$ as the product of the density $n_0 \mu/\bigl( V+ \sqrt{V^2 - \mu^2/2} \bigr)$ at the barrier boundary and the velocity $\hbar \ka/m$ at the barrier center.  
Alternatively, it is instructive to rewrite Eq. \ref{eq:jc} in terms of the bulk current density $c n_0$, and the tunneling amplitude $t_0(\tVb, d)$ across the barrier as
\begin{align}\label{eq:jc2}
j_c = c n_0 t_0(\tVb, d),
\end{align}
with
\begin{align}\label{eq:t0}
t_0(\tVb ,d) &=  2\sqrt{2} \frac{\sqrt{ 6\tVb -4 - \sqrt{4 \tVb^2 -2} } }{2 \tVb + \sqrt{4\tVb^2-2}}  \nonumber \\
&\quad \times \exp \Bigl(-\frac{d}{2\xi} \sqrt{ 6\tVb -4 - \sqrt{4 \tVb^2 -2} }   \Bigr).
\end{align}
$c= \sqrt{\mu/m}$ is the sound velocity, $\xi$ the healing length, and $\tVb = V/\mu$ the scaled barrier height. 
We note that $j_c$ is described in terms of the bulk current and the tunneling amplitude for a 3D condensate in Refs.  \cite{Zwerger1, Zaccanti}.
We determine the estimate $I_c= j_c Ly$ by using  $V=V_0$ and $d \approx  1.2 w$,  and by determining $c(T)$ and $n_0$ numerically. 
$V_0$ and $w$ are of the Gaussian barrier used in the simulation.
This value for $d$ is set by fitting the simulated critical currents in the IJJ regime, which is close to our assumption $d \approx w$ used in Ref. \cite{Niclas}.
In Figs. \ref{Fig:Ic}(b)-(d) we show the estimates  as a function of $\tV$ for $T/T_0=0.1$, $0.2$, and $0.3$.
The estimates agree with simulated critical currents at all $\tV > 1$, for all $T/T_0$. 
The agreement is particularly good for the IJJ regime. 
This suggests that the barrier reduction of $I_c$ is due to the tunneling amplitude $t_0(\tV, w)$ that decreases with increasing $\tV$ and $w$. We note that the barrier width reduction follows an exponential behavior given by Eq. \ref{eq:jc}.

\begin{figure}[]
\includegraphics[width=1.00\linewidth]{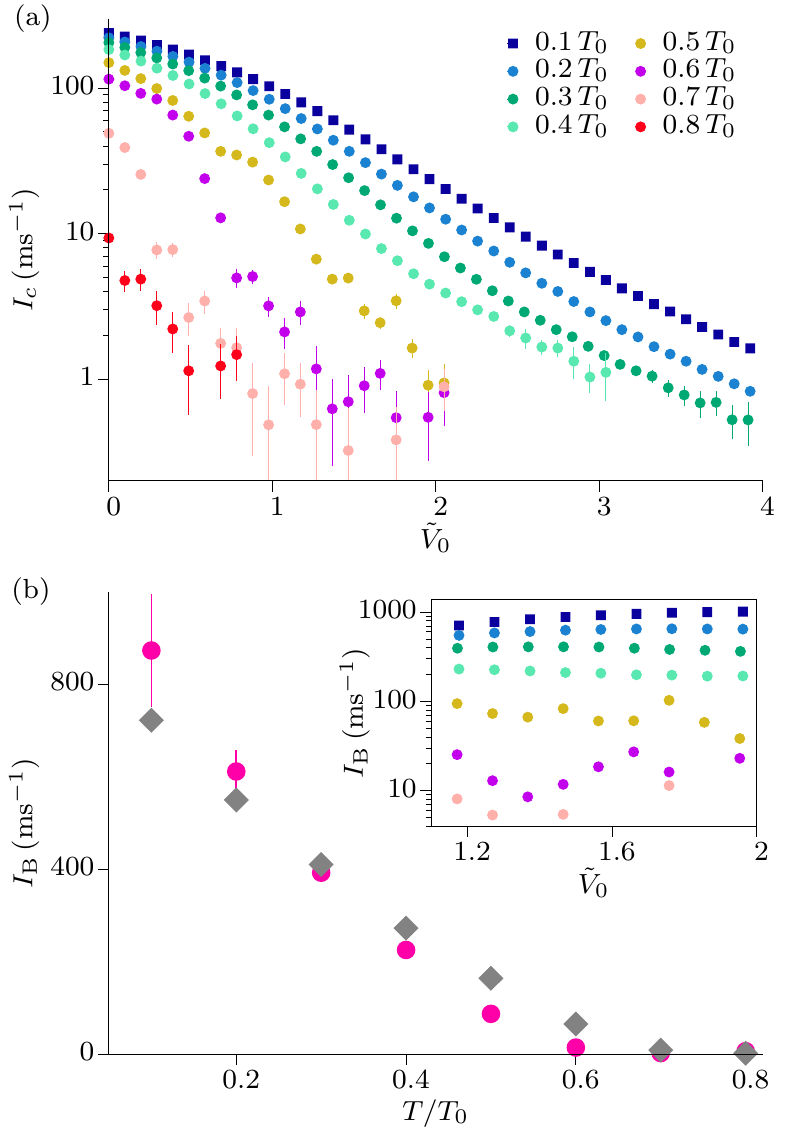}
\caption{\textbf{Temperature dependence.} 
(a) $I_c (\tV)$ for $w/\xi=2.9$ and various $T/T_0$.  
(b) Mean bulk current $I_\mB$ (circles) determined from the normalized critical currents shown in the inset, see text.
The error bar denotes the standard deviation. 
The bulk values of the current determined using the condensate density and phonon velocity are shown by the diamonds. 
}
\label{Fig:ic_temp}
\end{figure}

\begin{figure}[t]
\includegraphics[width=1.00\linewidth]{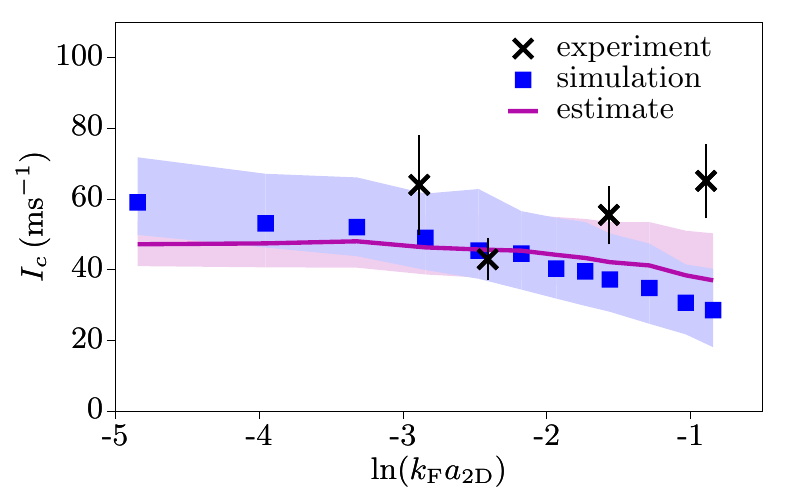}
\caption{\textbf{Comparison to the experiments.} The measurements of the critical current (crosses) are compared to the simulations (squares) and the analytical estimate (continuous line) for various $\ln(k_\mF a_\m2D )$ in the BEC regime. 
$k_\mF$ is the Fermi wavevector and $a_\m2D$ is the 2D scattering length.
We use $n \approx 1.24\, \mum^{-2}$, $T/T_0 \approx 0.3$, $w=0.81\, \mum$, and $\tV=1.4$, which are the same as the experiments.  
The shaded areas include a $15\%$ uncertainty of $\tV$ as in experiment.  
The measurement data is from Ref. \cite{Niclas}.
}
\label{Fig:comp}
\end{figure}

   In Fig. \ref{Fig:ic_temp}(a) we show $I_c$ as a function of $\tV$ for $w/\xi =2.9$ and various $T/T_0$.
As $T/T_0$ increases, $I_c(\tV)$ decreases with a rather sudden jump to very low values for $T/T_0 > 0.6$. 
$I_c$ and the bulk current are connected according to Eq. \ref{eq:jc2} via $ I_\mB = I_c(\tV) / t_0(\tV, w)$. 
We thus divide $I_c(\tV)$ by $t_0(\tV, w)$, see inset of Fig. \ref{Fig:ic_temp}(b). 
The results are almost independent of $\tV$ as expected. 
By taking an average over the range $\tV =1-2$, we obtain the mean value of $I_\mB$ which is shown in Fig. \ref{Fig:ic_temp}(b). 
The mean $I_\mB$ decreases with increasing $T/T_0$ and becomes small due to increased thermal fluctuations for $T/T_0 \geq 0.6$.
We compare this result to the actual bulk value $I_\mB = c(T) n_0 L_y$ by determining $n_0$ and $c(T)$ numerically for various $T/T_0$. 
This bulk result agrees at intermediate temperatures, where the system is near the IJJ regime. 
At low temperatures, the system is in the MM regime, where the above estimate is not valid. 
The bulk current is linked to the BKT scaling exponent that we determine in Sec. \ref{sec:exp}.

    Finally, we compare the simulations to the measurements that are performed at several interaction strengths in the BEC regime \cite{Niclas}.  
We use the range $\tilde{g}=1 - 4$, and determine the critical current $I_c$ as described above.  
In Fig. \ref{Fig:comp} we show the simulated $I_c$ as a function of the interaction parameter $\ln(k_\mF a_\m2D)$. 
$k_\mF$ is the Fermi wavevector and $a_\m2D =2.96 \ell_z \exp \bigl( -\ell_z\sqrt{\pi}/a_{\mathrm{3D}} \bigr)$ is the 2D scattering length, where $a_\mathrm{3D} = a_s$ is the 3D scattering length.  
The simulation results are consistent with the experimental results within the error bars of the measurement.
For $\ln(k_\mF a_\m2D)= -1$, the simulation results start to deviate from the experimental result, because the system enters the strongly interacting regime, where the c-field approach starts to deviate systematically.   
The shaded areas reflect the error bars of theory when assuming a $15\%$ uncertainty of $\tV$ as in experiment.  
We calculate the estimate of Eq. \ref{eq:jc} by determining the condensate density numerically for all interactions. 
The condensate fraction $n_0/n$ has weak interaction dependence and is about $62\%$. 
We show the analytical estimates in Fig. \ref{Fig:comp}.
The estimates, including the $15\%$ uncertainty of $\tV$, are consistent with both simulation results and the experimental results.

\section{Berezinskii-Kosterlitz-Thouless scaling}\label{sec:exp}
\begin{figure}
\includegraphics[width=1.00\linewidth]{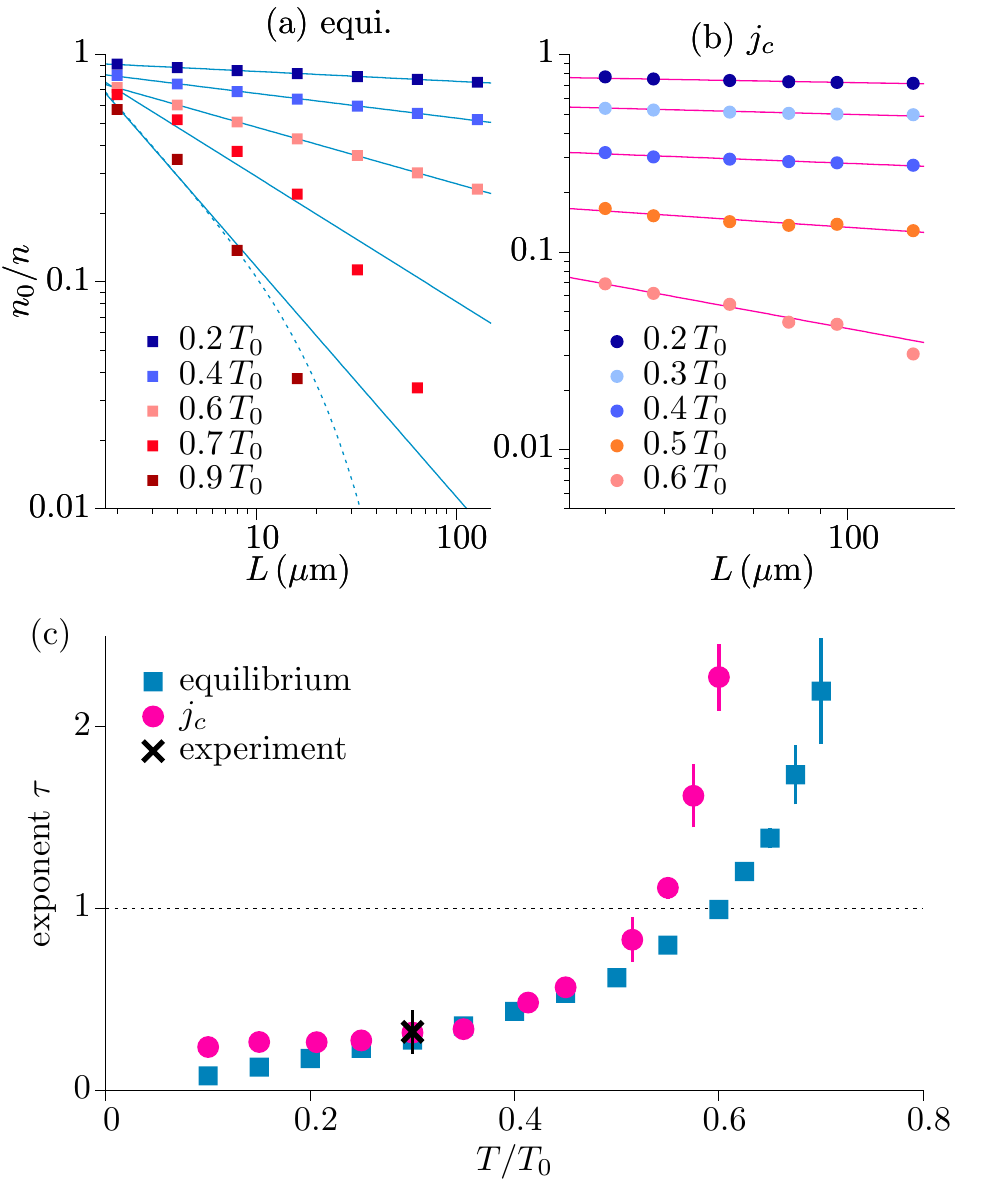}
\caption{\textbf{Determining the scaling exponent.} 
(a) Condensate fraction $n_0/n$ of the equilibrium system as a function of system length $L$ on a log-log scale for various $T/T_0$.
(b) Averaged $n_0/n$ determined from the critical current density $j_c$ via Eq. \ref{eq:jc2}.  
The continuous lines are the algebraic fits.
The exponential fit (dashed line) is a good fit for $T/T_0=0.9$ in panel (a).  
(c) Extracted exponents of the equilibrium system (squares) and $j_c$ (circles). 
The black cross corresponds to the measurements of $j_c$ in Ref. \cite{Niclas}.
  }
\label{Fig:tau}
\end{figure}

\begin{figure}[t]
\includegraphics[width=1.00\linewidth]{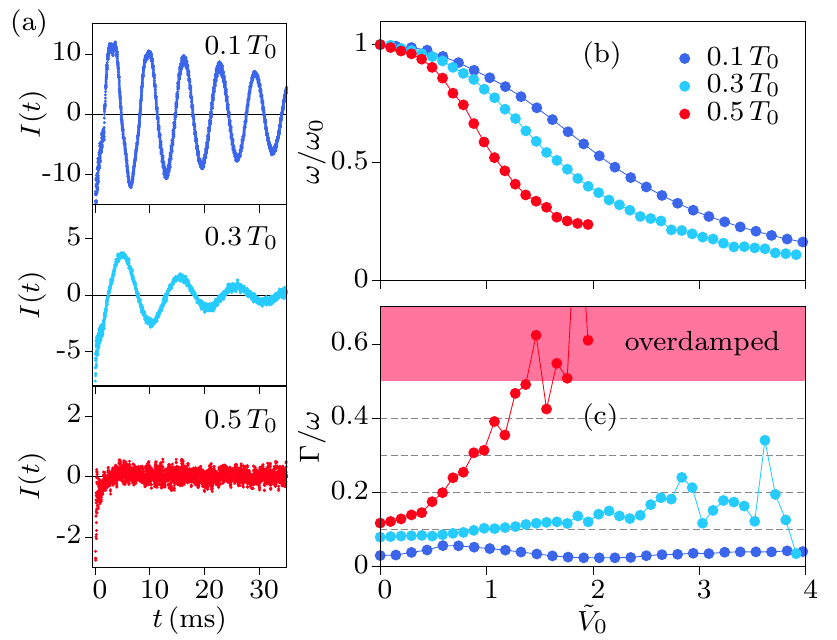}
\caption{\textbf{Current damping.} 
(a) Oscillations of the current $I(t)$ for $T/T_0=0.1$, $0.3$, and $0.5$. We use $\tV = 2$ and $w=1 \, \mum$.
(b) Normalized oscillation frequency $\omega/\omega_0$ and (c) scaled damping rate $\Gamma/\omega$, as a function of $\tV$, for the temperatures as in panel (a). 
$\omega_0$ is the sound frequency. 
The red shaded area denotes the overdamped regime.  
 }
\label{Fig:damping}
\end{figure}

\begin{figure*}
\includegraphics[width=1.00\linewidth]{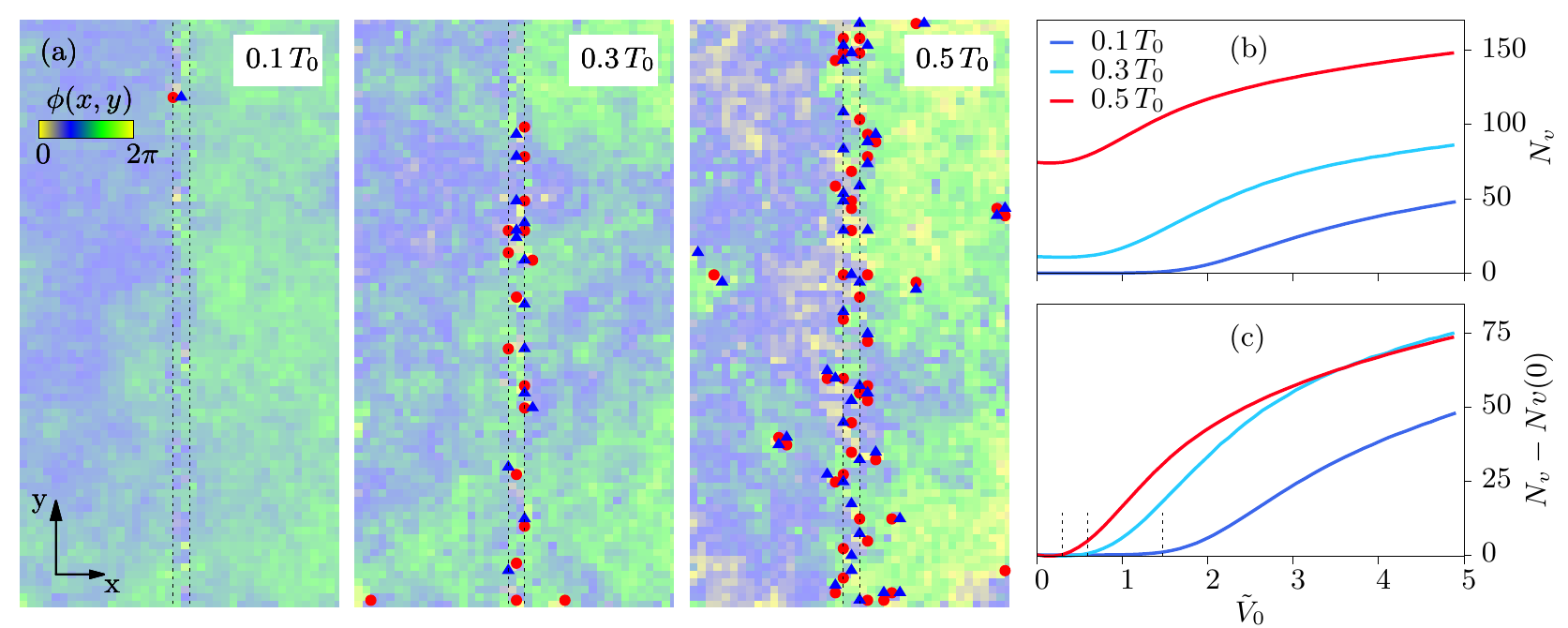}
\caption{\textbf{Damping mechanism.} 
(a) Phase distribution $\phi(x,y) = \phi(x,y) -  \phi_\mm$ of a single trajectory at $t=3.6\, \mms$, after the phase imprint, for $T/T_0=0.1$, $0.3$, and $0.5$. $ \phi_\mm$ is the mean global phase.
The corresponding bulk condensate fractions are $n_0/n=0.88$, $0.64$, and $0.40$, respectively.
The barrier parameters are the same as in Fig. \ref{Fig:damping}(a). The barrier width is indicated by the two vertical dotted lines.
The circles and the triangles denote vortices and antivortices, respectively. 
The box dimensions are $20 \times 40\,\mum^2$.   
For the time evolution dynamics see Ref. \cite{note_1}. 
(b) Average vortex number $N_v$ and (c) differential vortex number $N_v - N_{v,0}$, depicted as a function of $\tV$, 
where $N_{v,0}$ is $N_v(\tV =0)$. 
The vertical dashed lines in panel (c) mark the onset of barrier-induced vortices for $T/T_0=0.1$, $0.3$, and $0.5$ at $\tVc =1.47$, $0.58$, and $0.29$, respectively.   }
\label{Fig:vortex}
\end{figure*}

  We have shown above that the critical current depends on the condensate density. We use that  dependence to  extract the scaling exponent of the quasi-condensate.
The condensate density scales algebraically with the system size as 
\begin{align}\label{eq:n0}
n_0 \approx n \Bigl( \frac{L}{r_0}  \Bigr)^{-\tau/4}, 
\end{align}
where $\tau (T)$ is the temperature-dependent exponent, $L$ the system length, and $r_0$ the short-range cutoff of the order of $\xi$. 
Employing this scaling, we first determine $\tau (T)$ for a square-shaped system in equilibrium. 
We choose $n \approx 2.25\, \mum^{-2}$, $\tilde{g}=1.8$, and $L$ in the range $(2-128)\, \mum$. 
We calculate $n_0$ as a function of $L$ for various $T/T_0$ for a system with periodic boundary conditions.
In Fig. \ref{Fig:tau}(a) we show the condensate fraction $n_0/n$ as a function of $L$ for various $T/T_0$. 
At low and intermediate $T/T_0$, $n_0$ shows a power-law behavior as it decreases linearly with $L$ on a log-log scale. 
At high $T/T_0$, $n_0$ deviates from this power-law scaling and instead shows an exponential  behavior 
which is characteristic of the thermal phase \cite{Singh2014}. 
To confirm the power-law scaling, we fit $n_0/n$ to Eq. \ref{eq:n0}, with $\tau$ and $r_0$ as fitting parameters.
We show the fits in Fig. \ref{Fig:tau}(a). 
The algebraic fits describe the behavior very well for $T/T_0 \leq 0.6$, whereas they fail to capture the dependence for $T/T_0 > 0.6$. 
For $T/T_0=0.9$ the dependence is captured by the exponential fit, while for $T/T_0=0.7$ the algebraic fit is better than the exponential fit.

   Next, we determine the scaling exponent of the critical current density.
We choose $n$ and $\tilde{g}$ as above, and $L$ in the range $(40-128)\, \mum$.
For the barrier, we use $w/\xi= 2.9$ and $\tV$ in the range $\tV=1.2-2$. 
We calculate the critical current density $j_c$ as a function of $L$ for various $T/T_0$, 
where $j_c$ is determined by the critical current described in Sec. \ref{sec:jc}. 
We obtain the condensate density $n_0$ by normalizing $j_c$ with the sound velocity $c(T)$ and the tunneling amplitude $t_0(\tV, w)$, see Eq. \ref{eq:jc2}.
We then average $n_0$ over the barrier heights employed. 
We expand on this normalization in Appendix \ref{ap:sec:scale}.
In Fig. \ref{Fig:tau}(b) we show the condensate fraction $n_0/n$ that is determined from the critical current density as a function of $L$ for various $T/T_0$. 
$n_0/n$ shows a power-law behavior for $T/T_0 \leq 0.6$, 
which is confirmed by the algebraic fits shown in Fig. \ref{Fig:tau}(b). 
   In Fig. \ref{Fig:tau}(c) we show the temperature dependence of  $\tau$ for the equilibrium system and $\tau$ determined from the critical current density.  
 The equilibrium value of  $\tau$ increases linearly at low and intermediate $T/T_0$, 
while it deviates from linear behavior at high $T/T_0$ in the crossover regime. 
We estimate the transition temperature with the BKT critical value $\tau_c=1$, which gives $T/T_0 \approx 0.6$. 
We note that this value of the critical temperature is renormalized to a lower value in thermodynamic limit \cite{Giorgini2008}.  
We also note that this estimate is below $T_0$ of a weakly interacting 2D Bose gas \cite{Prokofev2001}.
Above the transition, $\tau$ increases rapidly with the temperature. 
This temperature dependence of $\tau$ across the transition is described by the RG equations \cite{LM2017}.
Studies of BKT scaling in ultracold gases were reported in Refs. \cite{Hadzibabic2006, Murthy2015, Igor2016}.
In addition to the equilibrium value of $\tau$ we show the value of $\tau$ based on the critical current scaling. 
The results show excellent agreement with the exponents of the equilibrium system for the temperatures $0.2 < T/T_0 < 0.55$, 
and follow the qualitative behavior outside of this temperature range. 
The deviations below $T/T_0=0.2$ are due to multimode dynamics that influence the results of the critical current density, while the deviations for $T/T_0 \geq 0.55$ are due to thermal excitations at the barrier, and the onset of overdamped dynamics.

     Furthermore, we determine the exponent $\tau$ from the measurements of the critical currents shown in Fig. \ref{Fig:comp}. 
Making use of  Eq.  \ref{eq:jc2}, these measurements yield the condensate fraction $n_0/n = 0.72 (8)$.  
For the system size $L$ in the experiment and  $r_0= \xi$, the scaling of Eq. \ref{eq:n0} results in an exponent of $\tau= 0.32 (12)$. 
We show this value of $\tau$ for $T/T_0=0.3$ in Fig.  \ref{Fig:tau}(c), which agrees with the exponents of the simulated critical current density and the equilibrium system.

\section{Current damping and dissipation mechanism}\label{sec:damping}

  Here we analyze the damping of the supercurrent and identify the associated dissipation mechanism.
As an illustration, we choose $w/\xi=2.9$ and $\tV=2.0$, and calculate the time evolution of the current $I(t)$ as described above. 
In Fig. \ref{Fig:damping}(a) we show $I(t)$ for $T/T_0=0.1$, $0.3$, and $0.5$.
The current oscillations are underdamped at $T/T_0=0.1$ and $0.3$.
The damping increases with increasing $T/T_0$.
For $T/T_0=0.5$,  the current undergoes an overdamped motion. 
To quantify this observation we determine the oscillation frequency $\omega$ and the damping rate $\Gamma$ as described in Sec. \ref{sec:method}. 
In Fig. \ref{Fig:damping}(b) we show $\omega/\omega_0$ determined as a function of $\tV$ for $T/T_0=0.1$, $0.3$, and $0.5$.
$\omega_0$ is the oscillation frequency for  $\tV=0$, which we refer to as the sound frequency. 
As $\tV$ increases, $\omega/\omega_0$ decreases.  This decrease is more pronounced for higher temperatures.
We note that the dependence of $\omega/\omega_0$ on the barrier height differs qualitatively between the low $(\tV < 1)$ and high $(\tV > 1)$ barrier regimes. 
In Fig. \ref{Fig:damping}(c) we show the results of $\Gamma/\omega$ as a function of $\tV$.
For $T/T_0=0.1$, $\Gamma/\omega$ is small at all $\tV$, confirming underdamped motion.  
For $T/T_0=0.3$, $\Gamma/\omega$ increases at high $\tV$ and is generally below $0.5$ that we use as the definition of the temperature-induced overdamped limit.  
For $T/T_0=0.5$, $\Gamma/\omega$ increases rapidly with $\tV$ and reaches the overdamped limit at $\tV \geq 1.5$.

    To identify the origin of the damping we examine the phase dynamics of a single trajectory of the ensemble. 
We calculate the phase $\phi(x,y)= \phi(x,y)-  \phi_\mm$ for the same parameters as in Fig. \ref{Fig:damping}(a), 
where $\phi_\mm$ is the mean global phase. 
In Fig. \ref{Fig:vortex}(a) we show $\phi(x,y)$ at $t =3.6 \, \mms$, after the phase imprint, for $T/T_0=0.1$, $0.3$, and $0.5$.
The phase imprint develops a phase difference between the two reservoirs and a corresponding phase gradient across the barrier. 
At low temperature the reservoir phase is weakly fluctuating as demonstrated by $\phi(x,y)$ at $T/T_0=0.1$.
The fluctuations of the phase increase with increasing $T/T_0$. 
The reservoir phase is moderately and strongly fluctuating for $T/T_0=0.3$ and $0.5$, respectively. 
This results in the creation of vortices, which is confirmed by the calculation of the phase winding around the plaquettes of the numerically introduced lattice. 
We calculate the phase winding around the lattice plaquette of size $l\times l$ using $\sum_{\Box} \delta \phi(x,y) = \delta_x\phi(x,y) + \delta_y\phi(x+l,y)+\delta_x\phi(x+l,y+l)+\delta_y\phi(x,y+l)$, 
where the phase differences between sites are taken to be $\delta_{x/y} \phi(x,y)  \in (-\pi, \pi]$. 
We show the calculated phase windings in Fig. \ref{Fig:vortex}(a). 
We identify a vortex and an antivortex by a phase winding of $2\pi$ and $-2\pi$, respectively. 
For $T/T_0=0.1$, we observe only one vortex-antivortex pair inside the barrier and no vortices in the bulk.
This scenario changes due to increased thermal fluctuations at high temperatures. 
For $T/T_0=0.3$, there is nucleation of multiple vortex pairs inside the barrier, and a few vortex pairs near the box edges. 
For $T/T_0=0.5$, we observe proliferating vortices in the regions of low densities around the barrier, and in the bulk.

  To understand the role of vortex fluctuations at the barrier,  we calculate the total number of vortices $N_v$ and average it over the thermal ensemble.    
In Fig. \ref{Fig:vortex}(b) we plot $N_v$ as a function of $\tV$ for the same values of $T/T_0$ as in Fig. \ref{Fig:vortex}(a).
At $T/T_0 =0.1$, $N_v$ remains close to zero for barrier heights below a threshold value and beyond this $N_v$ increases with increasing $\tV$.
At high temperatures the system features thermal vortices even in the absence of the barrier and the onset of barrier-induced vortices occurs at a $\tV$ lower than that at low temperature.  
To determine this threshold $\tVc$ we calculate the differential vortex number $N_v (\tV) = N_v - N_{v,0}$, where $N_{v,0} = N_v (\tV=0)$. 
We show $N_v (\tV)$ in Fig. \ref{Fig:vortex}(c).
We define $\tVc$ for which $N_v (\tV)$ approaches $1$. 
This gives $\tVc =1.47$, $0.58$, and $0.29$ for $T/T_0=0.1$, $0.3$, and $0.5$, respectively. 
This onset of vortices is associated with the damping of the oscillation shown above.

\section{Conclusions}\label{sec:con}

   In this paper, we have established a direct connection between the Josephson critical current and BKT scaling in an ultracold 2D Bose gas using classical field simulations.  
For this, we have examined the dynamics across a Josephson junction created by a tunnel barrier between two uniform 2D clouds of $^{6}\mathrm{Li}_2$ molecules, which is motivated by the experiments of Ref. \cite{Niclas}.
Based on the current-phase relation, we have mapped out the multimode, the second-harmonic (SH), the ideal junction (IJJ), and the overdamped regime as a function of the barrier height and the temperature. 
For the IJJ regime, we have derived an analytical estimate of the critical current, which is in good agreement with the simulations and the experiments \cite{Niclas}. 
We have demonstrated the BKT scaling of the critical current numerically by varying the system size. 
The scaling exponents of the critical current are in agreement with the exponents of the corresponding equilibrium system. 
Finally, we have addressed the damping of the current, which is due to phononic excitations in the bulk, and the nucleation of vortex pairs in the junction.

  In conclusion, we have discussed the  dynamics of atomic clouds in 2D, coupled via a Josephson junction, which  results in the hybridization of the bulk and tunneling dynamics. As such, it combines and relates two foundational effects of quantum physics, in particular condensation and Josephson oscillations. Our results demonstrate a method to measure a static property of  many-body order, in particular the condensate density, via a dynamical  oscillatory process, in particular Josephson oscillations. Both the principle of this method, as well as the presented discussion of dynamical regimes of this system, can be applied to a wide range of quantum gas systems, to gain insight into their dynamical and static properties.

\section*{acknowledgements}

We thank Markus Bohlen for his contributions on experimental work,  Thomas Lompe and Henning Moritz for their contributions during this joint work and careful reading of the manuscript, and Francesco Scazza and Alessio Recati for stimulating discussions. 
This work was supported by the European Union’s Seventh Framework Programme (FP7/2007-2013) under grant agreement No. 335431 and by the DFG in the framework of SFB 925 and the excellence clusters `The Hamburg Centre for Ultrafast Imaging’- EXC 1074 - project ID 194651731 and `Advanced Imaging of Matter’ - EXC 2056 - project ID 390715994.

\appendix

\section{Multimode versus Josephson regime}\label{ap:sec:cpr}

\begin{figure}[]
\includegraphics[width=1.00\linewidth]{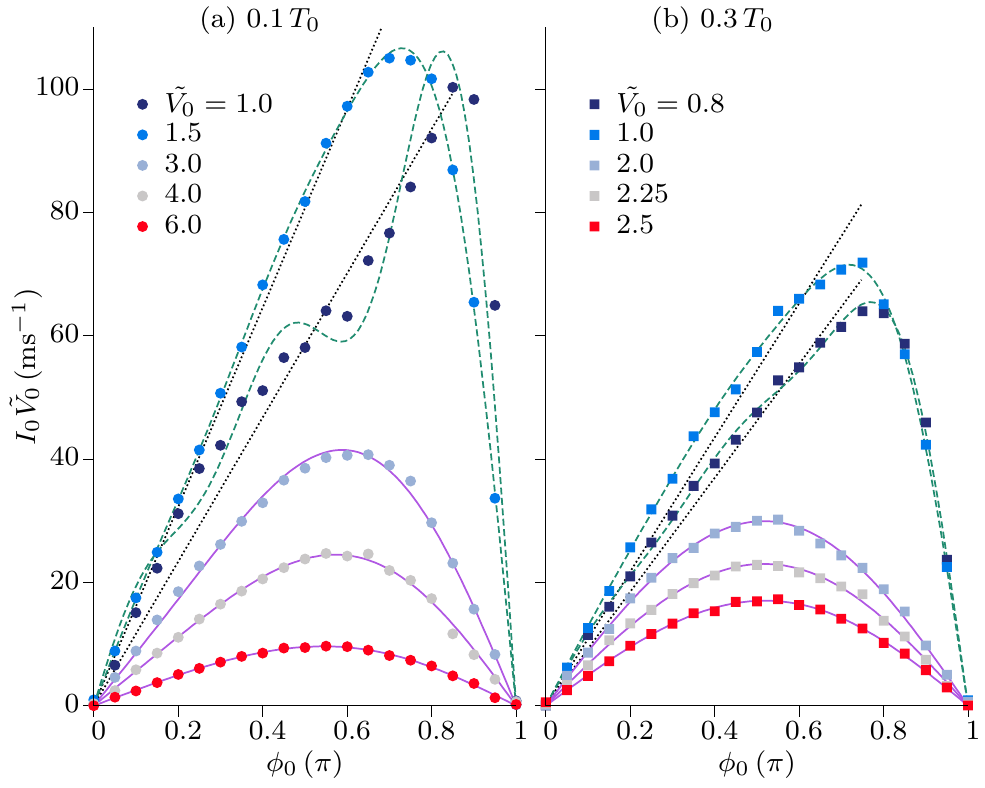}
\caption{\textbf{Transition between multimode  and Josephson regime.} 
$I_0 (\phi_0)$ for various $\tV$ at (a) $T/T_0=0.1$ and (b) $T/T_0=0.3$.  
The continuous and dashed lines are the second-harmonic and multi-harmonic fits,  respectively. 
The dotted lines are the linear fits.  }
\label{Fig:cpr}
\end{figure}

   In this appendix, we expand on the multimode regime of the current-phase relation (CPR). 
 We use the same parameters as in Sec. \ref{sec:bjj}, and calculate $I_0$ as a function of $\phi_0$ for various $\tV$. 
In Fig. \ref{Fig:cpr} we show these results at $T/T_0=0.1$ and $0.3$. 
As described in Sec.  \ref{sec:bjj}, we analyze these CPR curves by fitting them with a multi-harmonic fitting function  
$I( \phi_0) = \sum_{n=1}^{n_\mmax} I_n \sin(n  \phi_0)$, where we choose $n_\mmax=5$. 
At  $T/T_0=0.1$, the CPR curves are described by the multi-harmonic fits with $n_\mmax=5$ for $\tV  \leq 1.5 $, which we refer to as the multimode (MM) regime. 
For higher $ \tV $ we find the second-harmonic (SH) regime where $I_1$ and $I_2$ are non-negligible.
In contrast to the SH regime, the MM regime features a linear behavior up to a maximum value of the current for $\phi_0> \pi/2$. This is confirmed by the linear fits shown in  Fig. \ref{Fig:cpr}(a).
In Fig. \ref{Fig:cpr}(b) we show the CPR relations at $T/T_0=0.3$.
 For $\tV  \leq 1.0 $, the CPR curves display the MM regime which is also captured by the linear dependence up to a maximum value of the current for $\phi_0> \pi/2$.
For $\tV  \geq 2.0 $, the CPR reduces to the form of an ideal Josephson junction (IJJ), $I(\phi_0)= I_1 \sin \phi_0$, which we refer to as the IJJ regime.

\section{Estimate of the critical current}\label{ap:sec:Ic}

We derive an expression of the critical current by considering a rectangular barrier of width $d$ and height $V$ that is higher than the  mean-field energy $\mu$. 
We use the barrier ansatz  
\begin{equation}
\psi_l(x) = 
\begin{cases}
-\sqrt{n_0} \tanh \bigl( ( x + \delta)/( \sqrt{2} \xi ) \bigr) &  x < -d/2 \\
  A \exp \bigl( -\ka(x+d/2) \bigr) &  0 > x > -d/2
\end{cases}
\end{equation}
and 
\begin{align}
\psi_r(x) = 
\begin{cases}
A \exp \bigl( \ka(x - d/2) \bigr) &  d/2 > x > 0 \\
\sqrt{n_0} \tanh \bigl( ( x - \delta)/( \sqrt{2} \xi ) \bigr) &  x > d/2 
\end{cases}
\end{align}
$\psi_{l/r}$ are the fields of the left and right reservoir. $n_0$ is the density of the $k=0$ mode. 
$\delta$ and $\ka$ are determined by the continuity of the field and its derivative at $x= \pm d/2$. We include the mean-field repulsion in the barrier by determining $\ka$ variationally. The energy is 
\begin{align}
E= \frac{A^2}{2 \ka} \Bigl( \frac{\hbar^2 \ka^2}{2m} + V - \mu \Bigr) + \frac{g}{2} \frac{A^4}{4 \ka},
\end{align}
which we minimize using 
\begin{align} \label{Eq:ka}
\ka = \sqrt{k_0^2 + k_A^2}
\end{align}
with
\begin{align}
k_0^2 = \frac{2m(V-\mu)}{\hbar^2} \quad \text{and} \quad  k_A^2 = \frac{m g A^2}{2 \hbar^2}.
\end{align}
Continuity of wave function and its derivative, along with Eq. \ref{Eq:ka}, results in  
\begin{align}
\delta &= d/2 + \sqrt{2} \xi \arctanh(-A/\sqrt{n_0}), \\
A^2 &= n_0 \frac{\mu}{V + \sqrt{V^2 - \mu^2/2}}.  \label{Eq:A2}
\end{align}
We introduce a phase difference $\phi_j$ between the left and right reservoir as  $\psi(x) = \psi_l + \psi_r \exp(i \phi_j)$, and calculate the current density $j_x = \hbar/(2im) ( \psi^\ast \partial_x \psi -  \psi \partial_x  \psi^\ast )$ at $x=0$.
We find 
\begin{align}
j = 2 A^2  \frac{ \hbar \ka }{m} \exp(-\ka d)  \sin \phi_j. 
\end{align}
This is the current phase relation of a bosonic Josephson junction, with the critical current density 
\begin{align} \label{eq:ap:jc}
j_c &= 2 A^2  \frac{ \hbar \ka }{m} \exp(-\ka d).
\end{align}
$\ka$ and $A^2$ are given by Eqs. \ref{Eq:ka} and \ref{Eq:A2}, respectively. This result of $j_c$ is described in terms of the density $A^2$ at the barrier boundary and the velocity $\hbar \ka/m$ at the barrier center.

\section{Determining the scaling exponent }\label{ap:sec:scale}
\begin{figure}[]
\includegraphics[width=1.00\linewidth]{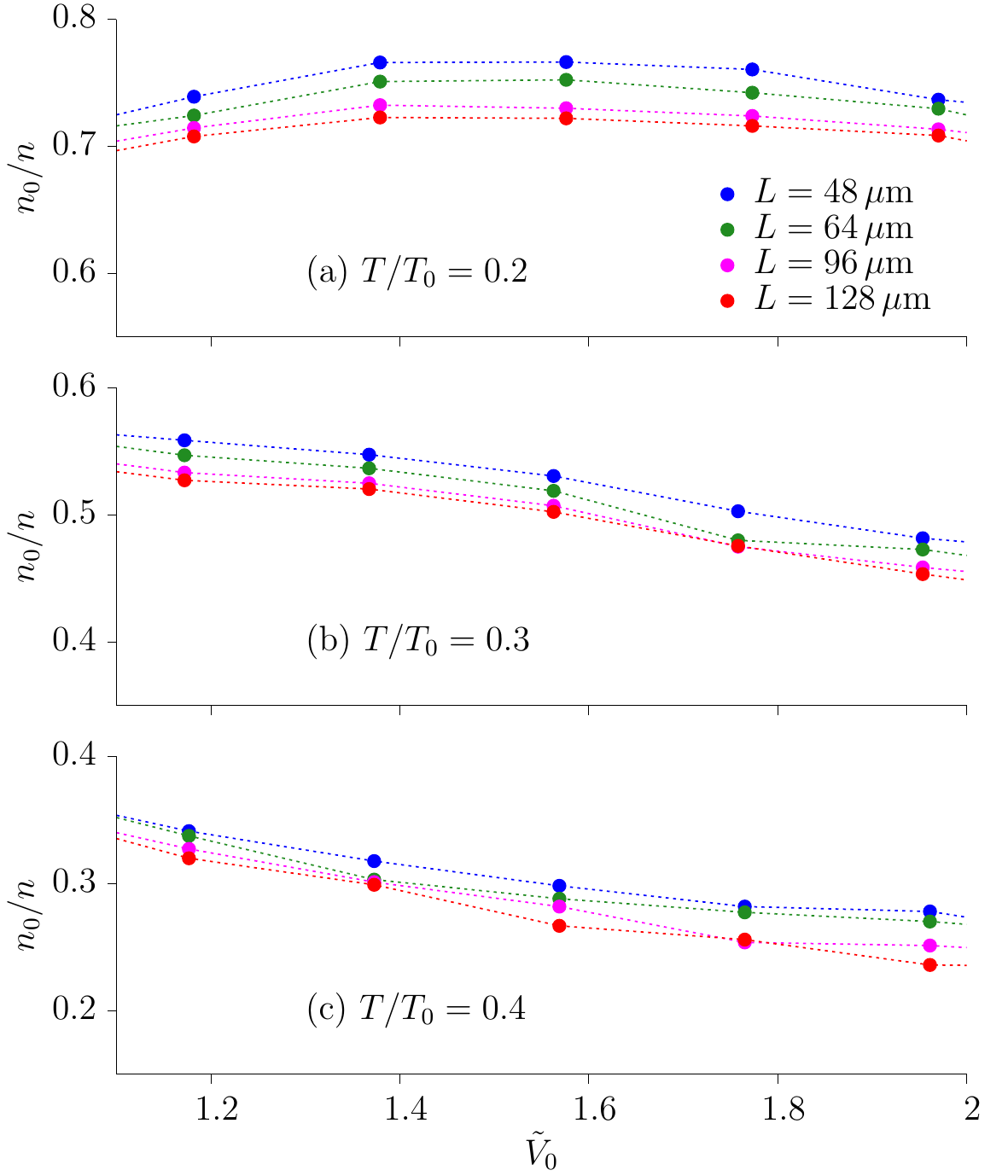}
\caption{Condensate fraction $n_0/n$ determined from the critical current density via Eq. \ref{eq:app:sc} for various values of the system length $L$ and $\tV$. 
We show $n_0/n$ for  $T/T_0=0.2$,  $0.3$, and $0.4$ in panels (a), (b), and (c), respectively. 
 }
\label{Fig:jc_n0}
\end{figure}
\begin{figure}[]
\includegraphics[width=1.00\linewidth]{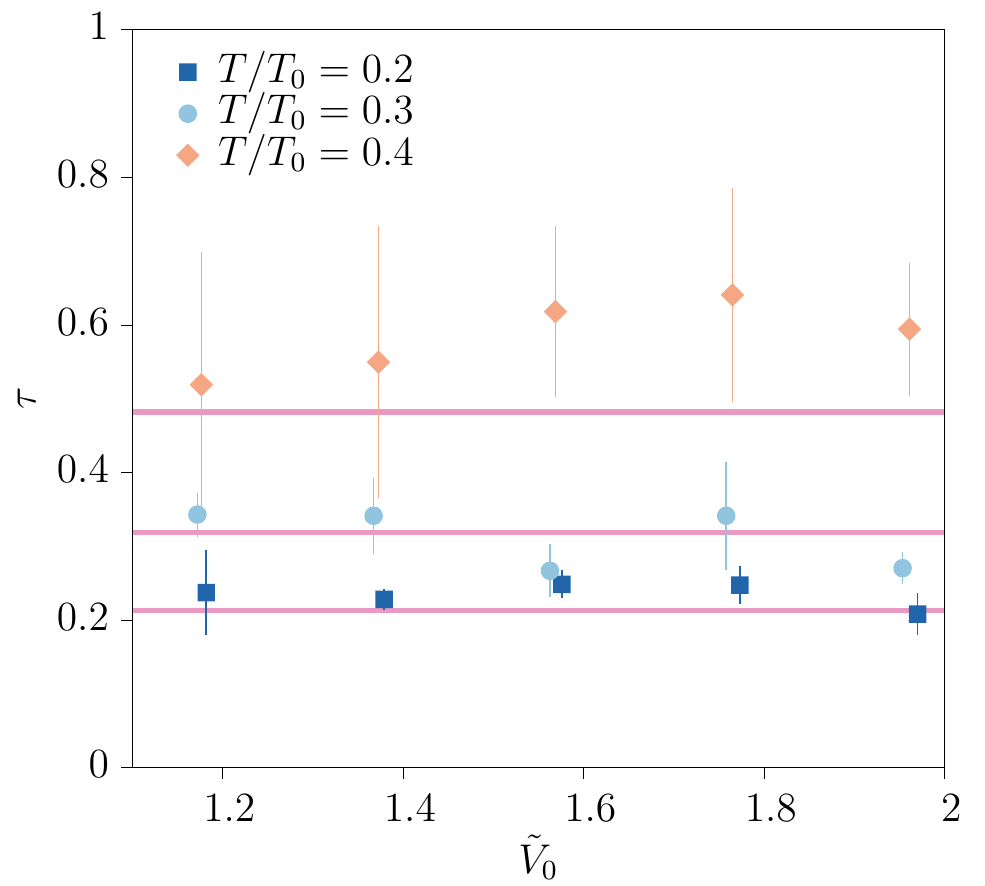}
\caption{Scaling exponent $\tau$ of the critical current density shown over the barrier height range $\tV=1.2 - 2$, for $T/T_0=0.2$, $0.3$, and $0.4$. 
$\tau$ is determined from the condensate fraction shown in Fig. \ref{Fig:jc_n0}. 
The horizontal lines depict the exponents determined from the condensate fraction that is averaged over the barrier heights employed.   
}
\label{Fig:tau_jc}
\end{figure}

We rewrite Eq. \ref{eq:ap:jc} as 
\begin{align}\label{eq:app:jc}
j_c = c n_0 t_0(\tVb, d),
\end{align}
with
\begin{align}
t_0(\tVb ,d) &=  2\sqrt{2} \frac{\sqrt{ 6\tVb -4 - \sqrt{4 \tVb^2 -2} } }{2 \tVb + \sqrt{4\tVb^2-2}}  \nonumber \\
&\quad \times \exp \Bigl(-\frac{d}{2\xi} \sqrt{ 6\tVb -4 - \sqrt{4 \tVb^2 -2} }   \Bigr).
\end{align}
$c= \sqrt{\mu/m}$ is the sound velocity, $\xi= \hbar/\sqrt{2\mu m}$ is the healing length, and $\tVb = V/\mu$ is the scaled strength.  From Eq. \ref{eq:app:jc}, the condensate density $n_0$ is 
\begin{align}\label{eq:app:sc}
n_0 = \frac{j_c}{c t_0(\tVb, d) }.
\end{align}

     To determine the algebraic scaling exponent of the quasi-condensate, we calculate $j_c$ as a function of $\tV$ for varying system sizes with simulations of the square-shaped box. 
We use  $w/\xi=2.9$, and $\tV$ in the range $1.2 - 2$.
The parameters $n$, $\tilde{g}$, and  $L$ are the same as in the main text in Sec. \ref{sec:exp}.
We determine $j_c$ as described in the main text. 
To obtain $n_0$ we divide $j_c$ by the sound velocity $c(T)$ and the tunneling amplitude $t_0(\tV, w)$, see Eq. \ref{eq:app:sc}.
In Fig. \ref{Fig:jc_n0} we show the condensate fraction $n_0/n$ as a function of $L$ for $T/T_0 = 0.2$, $0.3$, and $0.4$.
As expected, the results of $n_0/n$ are almost independent of $\tV$ and demonstrate a decreasing behavior with increasing $L$ and $T/T_0$. 
To determine the scaling exponent $\tau$, we fit $n_0/n$ to the function $n_0/n=(L/r_0)^{-\tau/4}$, with $\tau$ and $r_0$ as fitting parameters. 
We show the determined values of $\tau$ for $T/T_0 = 0.2$, $0.3$, and $0.4$ in Fig. \ref{Fig:tau_jc}.
The results are in agreement within the error bars for the barrier heights employed. 
For comparison, we average $n_0/n$ over the barrier heights employed and then determine $\tau$ from this averaged condensate fraction as we do in the main text in Sec. \ref{sec:exp}.
This result is also shown in Fig. \ref{Fig:tau_jc}, where it agrees with the determined exponents for $\tV$ in the range $1.2 - 2$.

\end{document}